# Extension of the Pierce model to multiple transmission lines interacting with an electron beam

Venkata Ananth Tamma and Filippo Capolino

Department of Electrical Engineering and Computer Science
University of California, Irvine, CA 92697, USA

*Abstract*: A possible route towards achieving high power microwave (HPM) devices is through the use of novel slow-wave structures, represented by multiple coupled transmission lines (MTLs), and whose behavior when coupled to electron beams has not been sufficiently explored. We present the extension of the one-dimensional linearized Pierce theory to MTLs coupled to a single electron beam. We develop multiple formalisms to calculate the *k*-ω dispersion relation of the system and find that the existence of a growing wave solution is always guaranteed if the electron propagation constant is larger than or equal to the largest propagation constant of the MTL system. We verify our findings with illustrative examples which bring to light unique properties of the system in which growing waves were found to exist within finite bands of the electron propagation constant and also present possible means to improve the gain. By treating the beam-MTL interaction as distributed dependent current generators in the MTL, we derive relations characterizing the power flux and energy exchange between the beam and the MTLs. For the growing wave, we show that the beam always behaves as an energy source causing power flux along the transmission lines. The theory developed in this paper is the basis for the possible use of degenerate band-edges in periodic MTL systems for HPM amplifiers.

**I. Introduction**

Ever since their introduction in the 1940's due to the pioneering work of Kompfner, Pierce and other co-workers, traveling wave tubes (TWT) have found immense applications as microwave and millimeter wave amplifiers in areas such as RADAR, wireless communications, satellite communications and electronic counter measure systems, among others. A review of the historical developments and progress achieved in the area of microwave vacuum electronic devices like the TWT is presented in [1]. A simplified and analytic theory of operation of the TWT was described by Pierce as a wave interaction phenomenon [2]-[6]. The one-dimensional Pierce theory represented the slow wave structure as an ideal one dimensional transmission line (TL) and modeled the electron beam as a one-dimensional "fluid". The amplification produced by the TWT was seen as the result of interaction between the RF signal on the TL with the traveling charge density waves on the electron beam. While the one-dimensional TL was described by the Telegrapher's equations, the electron beam was described by the equation of continuity and Newton's second law, with both equations expressed in Eulerian coordinates [2], [4], [6]. These inherently non-linear relations were then linearized under the assumptions of small signal amplitudes [2], [4]. Due to the linearization, the one-dimensional Pierce theory fails to account for the saturation of the TWT and does not model the inherent non-linearities in the TWT. However, despite the apparent oversimplification, the one-dimensional Pierce model has been used to analytically predict the performance of the TWT in the small signal regime with reasonable accuracy and has led to the accumulation of substantial physical insights into the operation of TWT-like devices.

In general, a slow wave structure supports many modes at a given frequency and therefore the sum of the fields due to these modes would interact with the electron beam. If, however, only one of the modes were to interact with the beam due to the synchronization of their phase velocities, then it would be justified to use the simplified single TL model as used in Pierce model to analyze the interaction between the beam and the RF fields. However, the exact form of the slow wave structure does not play a role in the Pierce model due to the very general nature of analysis

V. A. Tamma and F. Capolino, UC Irvine, May 2013

[6]. Some commonly used slow wave structures include the helix [6] and variants such as the ring-bar and the ring-loop structures [6]. Despite being a popular slow wave structure due to the inherent simplicity, an important fundamental drawback of the helix is its limited power handling capacity [6]. Recently, there has been an increased demand for high power microwave devices for both civilian and military applications [7] with the aim of achieving giga-watt peak power levels and very short pulse duration (order of tens of nanoseconds) microwave and millimeter-wave radiation. In light of this demand, it might be beneficial to explore unconventional slow wave structures with the aim of increasing the peak power levels and the operating efficiency of the TWT. A cursory analysis reveals that the single TL model does just well enough to reasonably represent the above slow wave structures at the interaction frequencies thereby justifying its use in the one-dimensional Pierce model. However, the representation of a slow wave structure as a single TL is one of Pierce theory's major limitations which can be overcome by use of MTLs, with each individual TL in the MTL representing a waveguide mode, including mutual coupling among TLs. Typically, slow wave structures are implemented as periodic circuits and therefore the operating frequency is chosen to be close to the band-edge of the dispersion diagram of the structure.

Previously, MTLs have been applied to TWTs both theoretically and experimentally from a few different perspectives. For example, a distinct TL (helix) was used to efficiently couple the RF signal in and out of the TWT helix at any point without making drastic modifications to the first TWT helix [8], [9]. While the coupled helices were well described using MTLs [8][9], it should be noted that in [8][9] (and references therein), the electron beam was typically assumed to be uncoupled with the external coupling helix (or TL). Also, Rowe has analyzed multiple interacting electron beam systems theoretically in terms of a coupled MTL system [10]. Furthermore, it was attempted to improve the interaction impedance and gain of a standard helix TWT by use of multiple helices [11], paralleled TWT circuits and electron beams [12] and multiple ladder circuits [13]. However, the theory developed in [11]-[13] is not general and does not delve into conditions for growing waves and energy conservation.

In the realm of periodic structures, recently, it has been proposed by Figotin and Vitebsky that a dramatic increase in field intensity and a vanishing group velocity associated with a degenerate band-edge (DBE) resonance could be achieved by use of specially arranged periodic arrays of anisotropic layers with mis-alignment of in-plane anisotropy between the anisotropic layers in a unit cell [14]-[17]. Previously, the DBE mode has been observed in periodic stacks of misaligned anisotropic layers [14]-[16], periodically loaded metallic waveguides [18] and coupled waveguiding systems [19], [20]. We anticipate that waveguiding structures exhibiting the DBE mode, modeled by MTLs, if suitably coupled to the electron beam, could be used to design TWT's with unprecedented peak power levels.

In this paper, we develop a general theory for the interaction of a MTL system with a single electron beam and obtain analytical results for the dispersion diagram of the combined beam-MTL system and derive conditions to obtain growing waves on the coupled MTL-beam system. We also analyze the power flow and energy conservation in the system and derive conditions for power transfer from the beam to the MTL. Sec. II-F presents illustrative examples for an electron beam coupled to a MTL comprising of multiple, identical, coupled TLs similar to the examples presented in [11]-[13] and shows an increase in gain similar to the conclusions of [11]-[13]. The theoretical formulation in this paper can be used to analyze MTL systems interacting with a single electron beam, multiple interacting modes in a single waveguide interacting with a single electron beam and it could be extended to the case of multiple interacting electron beams treated as a MTL system. Lastly, but importantly, the presented formulation is the first step toward the analysis MTLs exhibiting a DBE, coupled to an electron beam which may enable new ways to obtain large gains, and it will be the subject of future investigations.

In particular, here we present an analytical theory analyzing the interaction of an electron beam coupled to a lossless MTL system by extending the Pierce model to the case of MTL and retaining its simplicity. We assume a small signal approximation and neglect the effect of the space charge forces (those arising from a gradient of the charge density wave) on the longitudinal electric field, as was done in [4], [6]. Since MTLs can be used to model coupled systems with exotic dispersion characteristics, like for example, DBE modes, the extension of the Pierce model to a

V. A. Tamma and F. Capolino, UC Irvine, May 2013

MTL system could model the interaction of electron beams with DBE modes in coupled waveguiding systems with applications to HPM devices, to be summarized in future publications. In addition, the Pierce theory with MTLs could be applied to the study of an electron beam interacting with multiple modes in a single real waveguide or to possible novel designs having more than one waveguide interacting with the electron beam.

The paper is organized as follows: in Sec. II, we develop the theoretical framework for the electron beam and the MTLs and present multiple formalisms to calculate the $k$-$\omega$ dispersion relation including a matrix formulation and a formulation in terms of admittances more akin to the original Pierce derivation. In addition, we discuss the solutions to the dispersion relation with emphasis on identifying the conditions for achieving growing waves and study some illustrative examples. In Sec. III, we discuss the energy conservation relations for the system and analyze the direction of power flow for plausible solutions of the system.

## II. Formulations to derive $k$-$\omega$ dispersion relation

The system of $N$-TLs coupled among each other by inductive and/or capacitive means and to an electron beam along the $+z$ direction is shown in Fig. 1. The terms $s_n, a_n$, describe how the beam couples to the $N$-TLs and will be discussed later on in the paper.

### A. Electron beam dynamics

We follow the linearized equations which describe the space-charge wave on the electron beam presented by Pierce [4]. We assume a narrow cylindrical beam of electrons with constant axial electric field across the beam cross-section, and purely longitudinal electron motion and the absence of any transverse electric fields. Let the total electron-wave velocity along $+z$ direction be $u_0 + v$ [m/s], where, $u_0$ is the d.c. or average value and $v$ is the a.c. velocity modulation on the beam. We assume a line charge density with units [C/m] given by $\rho_0 + \rho_b$, where, $\rho_0$ is the d.c. or average value (assumed to be negative) and $\rho_b$ is the a.c. line charge density modulation on the beam. The total current with units of [A] flowing along the $+z$ direction is $I_0 + I_b$, where, $I_0$ is the d.c. or average value, and is negative since $I_0 = \rho_0 u_0$, whereas $I_b$ is the a.c. current modulation on the beam.

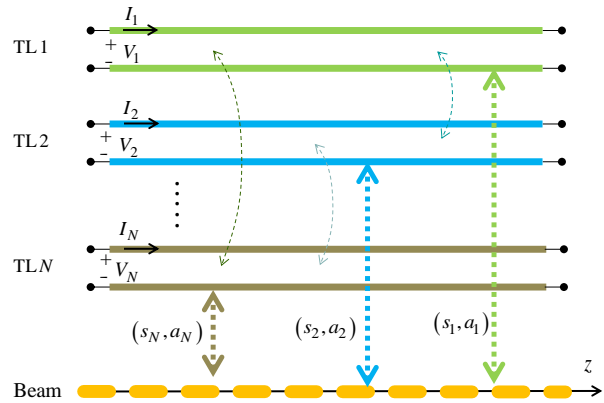

Fig. 1. Schematic describing the $N$-TLs coupled among each other and to the electron beam.

We denote the charge-to-mass ratio of the electron as $\eta = e/m$, where $e$ is the charge of the electron and is a negative number. The electronic propagation constant is defined as $\beta_0 = \omega/u_0$, where, $\omega$ is the radian frequency of operation in [rad/s]. The linearized equations of motion describing the electron beam are written as [4],

V. A. Tamma and F. Capolino, UC Irvine, May 2013

$$\partial_t V_b + u_0 \partial_z V_b = u_0 E_z, \tag{1}$$

$$I_b = \rho_b u_0 + \rho_0 v. \tag{2}$$

Here, we define an equivalent kinetic beam voltage $V_b = u_0 v/\eta$, with units of [V]. We assume that the axial electric field $E_z$ in (2) is the sum of the axial electric fields $E_{z,n}$ of all the $N$-TL modes,

$$E_z = E_{z,1} + E_{z,2} + .. + E_{z,N}. \tag{3}$$

It must be noted that effects of the space charge fields of the electron beam, modeled as plasma frequency reduction factor [6] have been neglected in this paper, as was done in [4], [6]. In addition to (1) and (2), the beam is further described by the one-dimensional continuity equation $\partial_t (\rho_0 + \rho_b) = -\partial_z (I_0 + I_b)$ that implies $\partial_t \rho_b = -\partial_z I_b$.

**B. Formulation of MTL with current generators**

In the Pierce model, the beam with current $I_0 + I_b$ was assumed to pass very close to the circuit, therefore inducing a current in the TL over the incremental distance $\Delta z$ [24], [25] and reflected in the TL equations as an equivalent parallel current generator per unit length, with unit [A/m] given by $i_S = -\partial_z I_b$. For $N$-TLs, we assume that the beam induces a current in every TL such that we consider parallel current generators on every TL $i_{S,n} = -\partial_z s_n I_b$ or, $\mathbf{i}_S = -\partial_z \mathbf{s} I_b$, and is schematically shown in Fig. 2. Here, $\mathbf{s} = [s_1, s_2, ..., s_N]^T$ is a column vector of dimensionless numbers called current interaction factors. Physically, the components of $\mathbf{s}$ scale the amount of charge induced on the waveguide walls by an electron beam and hence, denote the strength of each shunt current generator induced by the electron beam. In the original work of Pierce, the current interaction factor was a scalar number with unit value [4], [5]. Here, we include $\mathbf{s}$ to allow for different excitation of distinct TLs. We assume that the charge induced on the waveguide walls by the beam is instantaneous and hence, all elements of $\mathbf{s}$ are assumed to be real numbers.

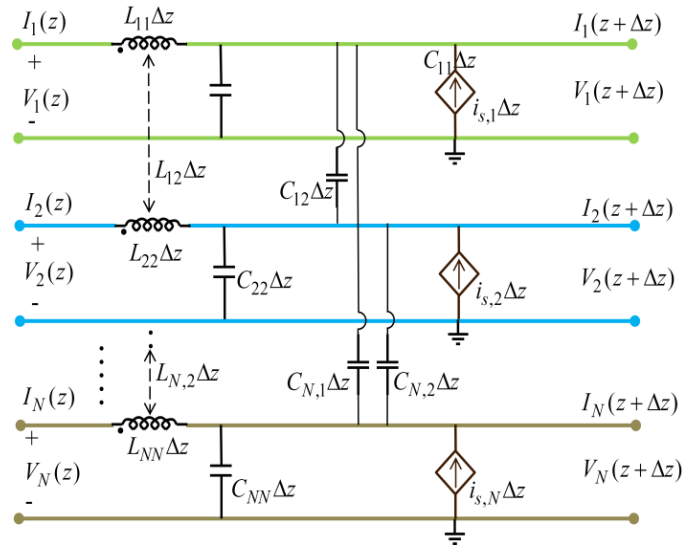

Fig. 2. Schematic showing the current generators per unit length $i_{S,n}$, one each per TL, representing the beam-TL interaction. The TL is schematically shown in terms of its per-unit-length inductances and capacitances along with the coupling inductances and capacitances.

V. A. Tamma and F. Capolino, UC Irvine, May 2013

The theoretical formulations for MTLs are well known and the approach presented in [21]-[23] is followed. For the MTL system with parallel current generators, we write the coupled equations as $\partial_z \mathbf{V} = -\underline{\mathbf{L}} \partial_t \mathbf{I}$ and $\partial_z \mathbf{I} = -\underline{\mathbf{C}} \partial_t \mathbf{V} + \mathbf{i}_S$. The TLs are defined by their real, distributed (per-unit-length) inductances ($L_{lm}$) and capacitances ($C_{lm}$), respectively. If $l = m$, then $L_{ll}$ and $C_{ll}$ represent the self-inductance and self-capacitance of the line $l$, and if $l \neq m$, then $L_{lm}$ and $C_{lm}$ represent the mutual or coupling inductances and capacitances between lines $l$ and $m$. We note that the matrices $\underline{\mathbf{L}}$ and $\underline{\mathbf{C}}$ are real symmetric and positive-definite [21]-[23]. Also, $\mathbf{V} = [V_1, V_2, ..., V_N]^T$ and $\mathbf{I} = [I_1, I_2, ..., I_N]^T$ are the voltages and currents on the TLs. We recall that a MTL system accurately represents wave propagation in a multimode real waveguide.

Considering the voltages and currents on the MTLs to vary with time and distance as $e^{j\omega t} e^{-jkz}$, where $k$ is the complex propagation constant, we rewrite the coupled equations describing MTL in frequency domain as

$$\partial_z \mathbf{V} = -j\omega \underline{\mathbf{L}} \mathbf{I}, \tag{4}$$

$$\partial_z \mathbf{I} = -j\omega \underline{\mathbf{C}} \mathbf{V} + \mathbf{i}_S. \tag{5}$$

Equations (4) and (5) are combined into a single second order differential equation in $\mathbf{V}$ given by $\left( \partial_z^2 + \omega^2 \underline{\mathbf{L}} \underline{\mathbf{C}} \right) \mathbf{V} = -j\omega \underline{\mathbf{L}} \mathbf{i}_S$. Assuming wave solutions, the second order differential equation is written as

$$\underline{\mathbf{Y}} \mathbf{V} = \mathbf{i}_S, \tag{6}$$

where, the circuit admittance *per unit length* [Ω/m] $\underline{\mathbf{Y}}$ is defined as

$$\underline{\mathbf{Y}} = \frac{1}{j\omega} \underline{\mathbf{L}}^{-1} \left( k^2 \underline{\mathbf{1}} - \omega^2 \underline{\mathbf{L}} \underline{\mathbf{C}} \right) \tag{7}$$

and $\underline{\mathbf{1}}$ stands for an identity matrix of size $N \times N$. For a single TL, Pierce assumed that the axial electric field was related to the TL voltage as $E_z = -\partial_z V$ [4], [5]. For MTLs, we assume that the axial electric field of the $n^{th}$ TL is related to only the voltage on the $n^{th}$ TL as $E_{z,n} = -a_n \partial_z V_n$ such that the total axial electric field is

$$E_z = -\sum_{n=1}^{N} a_n \partial_z V_n = -\partial_z \left( \mathbf{a}^T \mathbf{V} \right), \tag{8}$$

where, $\mathbf{a}^T = [a_1, a_2, ..., a_N]$ is a vector of dimensionless real numbers called field interaction factors. Physically, the components of $\mathbf{a}$ describe how much each individual TL affects the electron beam dynamics. In the original work of Pierce with a single TL, the field interaction factor was a scalar number with unit value [4], [5].

### C. Derivation of *k*-ω dispersion relation using impedance formulation

We derive a *k*-ω dispersion relation using an impedance formulation on the lines of the derivation of the fourth order equation using impedance concepts by Pierce [4], [5]. It is well known that dependent sources have been used to describe gain in transistors and linear amplifiers and therefore we find this point of view advantageous even here. The electron beam interaction with the TL system could be completely modeled as an active TL with a voltage dependent current source, shown schematically in Fig. 2, given by

V. A. Tamma and F. Capolino, UC Irvine, May 2013

$$\mathbf{i}_S = jk\mathbf{s}I_b = -\underline{\mathbf{Y}}_b\mathbf{V}, \tag{9}$$

where, the electronic beam admittance *per unit length* [Ω/m] $\underline{\mathbf{Y}}_b$ is defined as

$$\underline{\mathbf{Y}}_b = Y_b\,\mathbf{sa}^T, \quad Y_b = -j\frac{\eta\rho_0\beta_0 k^2}{u_0(\beta_0 - k)^2}. \tag{10}$$

(Note $Y_b = k^2 Y_e$ where $Y_e$ is the impedance defined in [4], [5].) In other words, thanks to the beam-MTL interaction, the generators can be seen as active admittances as shown in Fig. 2. The beam admittance per unit length $\underline{\mathbf{Y}}_b$, is derived from the second order differential equation obtained by combining (1) and (2) and also using (8) for solutions varying with time and distance as $e^{j\omega t}e^{-jkz}$. Equating (6) and (9), the resonance condition for MTLs interacting with an electron beam is

$$(\underline{\mathbf{Y}} + \underline{\mathbf{Y}}_b)\mathbf{V} = 0, \tag{11}$$

from which we obtain the k-ω dispersion relation by imposing $Det(\underline{\mathbf{Y}} + \underline{\mathbf{Y}}_b) = 0$ as

$$Det\left((k^2\underline{\mathbf{1}} - \omega^2\underline{\mathbf{LC}}) + \frac{\eta\rho_0\beta_0^2 k^2}{(k-\beta_0)^2}\underline{\mathbf{L}}\mathbf{sa}^T\right) = 0. \tag{12}$$

In fact, (12) can be reduced to a polynomial equation of order $(2N+2)$. This is shown by use of the fact that for any invertible $(N \times N)$ matrix $\underline{\mathbf{X}}$, any $(N \times 1)$ vector $\mathbf{A}$ and any $(1 \times N)$ vector $\mathbf{B}$, one has $Det[\underline{\mathbf{X}} + \mathbf{AB}] = Det[\underline{\mathbf{X}}][1 + \mathbf{B}\underline{\mathbf{X}}^{-1}\mathbf{A}]$ [26], leading to

$$D(k,\omega)\,P(k,\omega) = 0, \tag{13}$$

where, $D(k,\omega) = Det(k^2\underline{\mathbf{1}} - \omega^2\underline{\mathbf{LC}})$ and $P(k,\omega) = 1 + k^2\beta_0^2\eta\rho_0(k-\beta_0)^{-2}\mathbf{a}^T(k^2\underline{\mathbf{1}} - \omega^2\underline{\mathbf{LC}})^{-1}\underline{\mathbf{L}}\mathbf{s}$ are polynomials in *k* of degree 2*N* and 2*N*+2, respectively. The impedance formulation using dependent current sources can be further explored based on the schematic in Fig. 3 in which the transverse resonance condition is understood from a circuit point of view. We assume that the beam and MTL circuit quantities are interlinked by a transformer as follows. We note that the relation between primary and secondary currents and voltages are $i_{S,n} = s_n(jkI_b)$ and $-jkI_b = Y_b\sum_{n=1}^{N} a_n V_n$ in Fig. 3, when $s_n = a_n$, i.e., when $\mathbf{s} = \mathbf{a}$, which is what has been traditionally assumed for a single TL interacting with an electron beam [3]-[6]. The circuit also illustrates how each equivalent dependent current generator $i_{S,n}$ depends on the interaction of the beam with the other TLs via the beam impedance per unit length $Y_b$. It also offers other avenues to find and interpret the resonance condition and power relations in this MTL system. It will be clear that when the beam transfers power to the MTL system, the real part of $Y_b$ is negative, and it is seen to be the only circuit element able to provide power. By inspecting (10) one can observe that power transfer occurs only if *k* is complex.

V. A. Tamma and F. Capolino, UC Irvine, May 2013

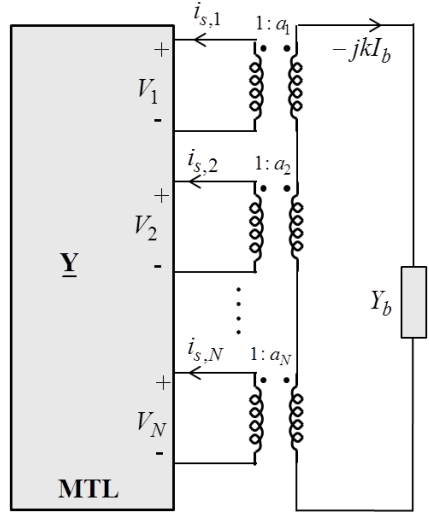

Fig. 3. Schematic showing the transverse resonance condition from a circuit point of view.

**D. Derivation of *k*-ω dispersion relation using a transfer matrix formulation**

The first order differential equations describing the beam and MTLs can be combined into a useful transfer matrix formulation. Assuming wave solutions, (1) is written in frequency domain as $\partial_z V_b = -j\beta_0 V_b + E_z$ and using (8) and (4), we obtain

$$\partial_z V_b = -j\beta_0 V_b + j\omega \mathbf{a}^T \underline{\mathbf{L}} \mathbf{I}. \tag{14}$$

Similarly, (2) is written in frequency domain as $\partial_z I_b = -j\beta_0 I_b + j\omega\eta(\rho_0/u_0^2)V_b$ and when combined with (5) gives

$$\partial_z \mathbf{I} = -j\omega\underline{\mathbf{C}}\mathbf{V} + j\beta_0 \mathbf{s} I_b - j\omega\eta \frac{\rho_0}{u_0^2} \mathbf{s} V_b. \tag{15}$$

Collecting the frequency domain forms of (1) and (2) along with (14) and (15), the coupled beam-MTL system could be represented by the set of (2*N*+2) first-order equations written in condensed matrix form

$$\partial_z \boldsymbol{\psi}(z) = -j\underline{\mathbf{M}}\,\boldsymbol{\psi}(z), \tag{16}$$

where, we define the state-vector $\boldsymbol{\psi}(z) = [\mathbf{V}\ \ \mathbf{I}\ \ V_b\ \ I_b]^T$. The matrix $\underline{\mathbf{M}}$ of size $(2N+2)\times(2N+2)$ is

$$\underline{\mathbf{M}} = \begin{bmatrix} \mathbf{0} & \omega\underline{\mathbf{L}} & 0 & 0 \\ \omega\underline{\mathbf{C}} & \mathbf{0} & \omega\eta\dfrac{\rho_0}{u_0^2}\mathbf{s} & -\beta_0\mathbf{s} \\ \mathbf{0} & -\omega(\mathbf{a}^T\underline{\mathbf{L}}) & \beta_0 & 0 \\ 0 & 0 & -\omega\eta\dfrac{\rho_0}{u_0^2} & \beta_0 \end{bmatrix}. \tag{17}$$

We recognize (16) to be the well-known Cauchy problem and assuming known initial conditions $\psi(z_0) = \psi_0$, the solution of (16) is written as $\psi(z_1) = \underline{\mathbf{T}}(z_1, z_0)\psi(z_0)$, where, we define $\underline{\mathbf{T}}(z, z_0)$ as the transfer matrix which uniquely relates the state vector $\psi(z)$ between two known points $z_0$ and $z_1$ along the $z$ axis. The transfer matrix formulation is also useful to analyze periodic systems such as coupled transmission lines exhibiting a DBE condition and will be elaborated in future publications. The propagation constants of the coupled system could be obtained directly by calculating the eigenvalues of the matrix $\underline{\mathbf{M}}$. Or, the $k$-$\omega$ dispersion relation of the coupled system could be obtained directly by evaluating the characteristic equation $Det(\underline{\mathbf{M}} - k\mathbf{I}) = 0$ of the matrix $\underline{\mathbf{M}}$. For $N$-TLs, the number of roots of the characteristic equation and hence the number of propagation constants of the system is $(2N+2)$. Following the derivation in Appendix A, $Det(\underline{\mathbf{M}} - k\mathbf{I}) = 0$ can be reduced to the form given in (13) which leads to a polynomial equation of degree $2N+2$ in $k$, and thus yielding $2N+2$ roots. We note that both this formulation and the one in Sec. II-C yield the same $k$-$\omega$ dispersion relation. A form of (13) with $\underline{\mathbf{L}}, \underline{\mathbf{C}}$ matrices in diagonal form is presented in Appendix B.

### E. Solutions of the $k$-$\omega$ dispersion relation

Since the dispersion relation in (13) is a product of two functions, for any fixed $\omega$ the solutions of (13) contain the solutions obtained from the conditions $D(k, \omega) = 0$ and $P(k, \omega) = 0$. The condition $D(k, \omega) = 0$, yields the $2N$ natural propagation constants $\{\beta_{c,1}, \beta_{c,2}, ..., \beta_{c,N}\}$ of the MTL *uncoupled* to the electron beam. However, it is evident from (13) that the solutions of $D(k, \omega) = 0$ cannot be solutions of $P(k, \omega) = 0$. Following the brief proof presented in Appendix B, for the $N$-line system coupled to electron beam, the existence of a complex conjugate root, and hence the existence of increasing and decreasing waves, is always guaranteed if

$$\beta_0 \geq \max\{\beta_{c,1}, \beta_{c,2}, ..., \beta_{c,N}\}. \tag{18}$$

Due to (18), we can obtain a maximum of $2N$ purely real propagation constants $k = (\beta_1, \beta_2, \beta_3, ..., \beta_{2N})$ and at least one pair of complex conjugate solutions $k = \beta \pm j\alpha$. A more general mathematical proof of (18) can be found in [23]. We numerically explore (18) for some illustrative examples in Sec. II-F. From Fig. 4 and Fig. 5, a complex $k$ solution is always found for $\beta_0$ satisfying (18). Furthermore, in the range $\min\{\beta_{c,n}\} < \beta_0 < \max\{\beta_{c,n}\}$, complex conjugate roots are seen to exist only when $\beta_0$ is in the vicinity of $\beta_{c,n}$. We find that a resonant-like behavior in Im($k$) when $\beta_0 \to \beta_{c,n}$ in which Im($k$) is seen to peak when $\beta_0 \approx \beta_{c,n}$ and then decays for $\beta_0$ larger than $\beta_{c,n}$, eventually approaching zero. In general, complex roots could also be found when $\beta_0 < \min\{\beta_{c,n}\}$ under certain conditions, as shown in [29], where a graphical analysis to determine the roots is also provided.

### F. Illustrative examples showing condition for growing waves

The inequality (18) establishing the condition for growing waves is numerically investigated for $N=2$, and $N=3$ TL-beam systems. The parameters used in the numerical study are detailed in Appendix C. The propagation constants were obtained by evaluating the eigenvalues of (17). The real ($\beta$) and imaginary part ($\alpha$) of the six solutions of the propagation constant $k$ for the 2-TL system are plotted in Fig. 4, versus $\beta_0$ obtained by keeping $\omega$ constant and varying the d.c. electron velocity $u_0$. In Fig. 4(a), six real values of $k$ are plotted, of which four correspond to 2 forward ($\beta > 0$) and 2 backward ($\beta < 0$) waves with constant values equal to $+\beta_{c,1}, +\beta_{c,2}$ and $-\beta_{c,1}, -\beta_{c,2}$ respectively. The remaining two identical real values of $k$, seen to linearly increase with $\beta_0$, correspond to the

V. A. Tamma and F. Capolino, UC Irvine, May 2013

complex conjugate roots $\beta \pm j\alpha$, whose imaginary parts are plotted in Fig. 4(b). The inequality (18) is confirmed in Fig. 4(b) which shows that the Im($k$) is non zero for $\beta_0 \geq \max\{\beta_{c,1}, \beta_{c,2}\}$, where we assume $\beta_{c,2} > \beta_{c,1}$. In addition, we observe that Im($k$) is nonzero in the vicinity of $\beta_0 = \beta_{c,1}$. However, it is seen that the Im($k$) vanishes for a large range of $\beta_0$ when $\beta_{c,1} < \beta_0 < \beta_{c,2}$.

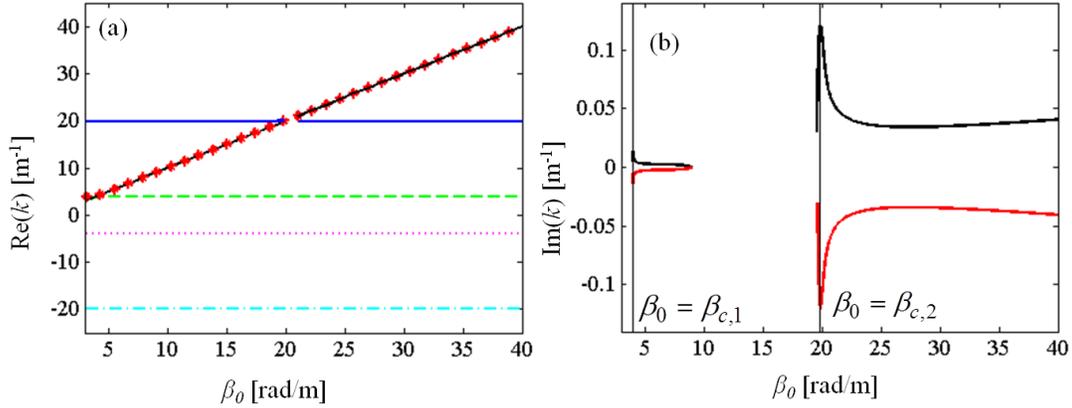

Fig. 4. Plots of (a) real ($\beta$) and (b) imaginary parts of the six solutions for $k$ for a 2-TL system coupled to the electron beam, versus $\beta_0$.

This behavior is further confirmed in Fig. 5(a) which plots the imaginary part of $k$ for the 3-TL system coupled to the electron beam. It is seen that Im($k$) has a continuous 'spectrum' when $\beta_0 \geq \max(\beta_{c,1}, \beta_{c,2}, .., \beta_{c,N})$ and in the range of $\beta_{c,1} \leq \beta_0 < \max(\beta_{c,1}, \beta_{c,2}, .., \beta_{c,N})$ it is seen that Im($k$) is non-zero within certain discrete bands which occur when $\beta_0 \rightarrow \beta_{c,n}$. In Fig. 5(b), we plot Im($k$) for the 3-TL system with all identical lines coupled to the electron beam, plotted in solid lines, which exhibits a similar behavior to that of a 1-TL system, plotted in dashed lines. The inductance and capacitance of the 1-TL was chosen to be the same as the self-inductance and self-capacitance of the 3-TL system. The coupling inductances and capacitances of the 3-TL system were chosen to be much smaller than the self-inductance and self-capacitances so as to enable a better comparison. The self-inductance of the line was chosen to be much larger than the self-capacitance so as to maximize the characteristic impedance of the line and hence the gain. From Fig. 5(b), we find that coupling the electron beam to a 3-TL system with all three identical lines improves the peak value of Im($k$) by about 45 % when compared to a 1-TL with same line parameters. This shows that in order to achieve the same peak value of Im($k$) as that of a 1-TL system, the line charge density, $\rho_0$, for the 3-TL system with three identical lines, could be reduced by as much as 67 % when compared to the 1-TL system. Otherwise for the same beam power, a 3-TL system achieves larger gain than with a single TL. We envision that by introducing periodic MTL systems operating near the Brillouin zone edge, the stored electromagnetic energy density in the MTL could be much higher than the corresponding density in 1-TL system.

V. A. Tamma and F. Capolino, UC Irvine, May 2013

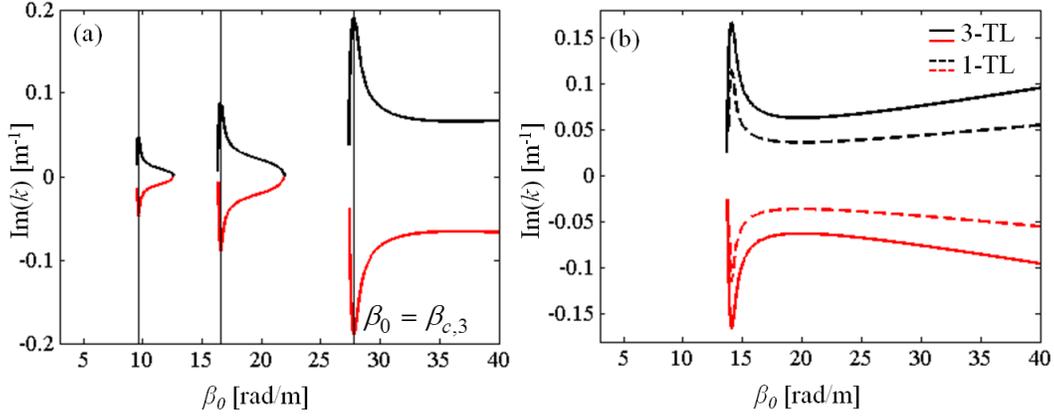

Fig. 5. Plots of imaginary part of $k$ for (a) 3-TL systems coupled to the electron beam, versus $\beta_0$. (b) 3-TL system with all three identical lines coupled to the electron beam, versus $\beta_0$. Also, plotted for comparison is the imaginary part of $k$ for a 1-TL system with parameters similar to the 3-TL system.

As an illustrative example, the effect of the variation of the coupling inductances and capacitances of a 2-TL system with all identical lines, coupled among themselves and to an electron beam, on Im($k$) is shown in Fig. 6. Here we plot the ratio of Im($k$) of the 2-TL system to the Im($k$) of a 1-TL system [ $\mathrm{Im}(k)_{2-TL} / \mathrm{Im}(k)_{1-TL}$ ] as a function of the variation in the normalized coupling inductance ($L_{12}/L_{11}$) and normalized coupling capacitance ($C_{12}/C_{11}$) of the 2-TL system with identical lines. Here, we denote $L_{12}$ and $C_{12}$ as the coupling inductance and capacitance, respectively, between the two TLs in the 2-TL system. Also, $L_{11}$ and $C_{11}$ denote the self inductance and capacitance, respectively, of both the identical TLs in the 2-TL system (i.e., $L_{11} = L_{22}$ and $C_{11} = C_{22}$). The Im($k$) of the 1-TL system was calculated by considering $L_{11}$ and $C_{11}$ as the self inductance and capacitance, respectively, of the 1-TL. A wide range of values of coupling inductance and capacitance were considered so as to include cases of both weak and strong coupling. The results in Fig. 6 clearly identify ranges of normalized coupling inductance and capacitance in which the Im($k$) of the 2-TL system can be either larger or smaller than the Im($k$) of the 1-TL system. This suggesting that the use of suitable operating parameters so that the 2-TL system coupled to an electron beam has an improved Im($k$) than the 1-TL system coupled to an electron beam. For all calculations related to the 2-TL system, we assumed $\beta_0 = \max\{\beta_{c,1}, \beta_{c,2}\}$ to always ensure the existence of growing waves on the system. Additionally, note that when $L_{12} = 0$ [H] and $C_{12} = 0$ [F] (corresponding to the origin in Fig. 6) we know that $\beta_0 = \beta_{c,1} = \beta_{c,2} = \beta_c$ where we denote as $\beta_c$ the propagation constant of the 1-TL; in this case we obtain $\mathrm{Im}(k)_{2-TL} = 1.26\,\mathrm{Im}(k)_{1-TL}$ showing that two identical TLs uncoupled with each other inductively or capacitively have Im($k$) larger than $\mathrm{Im}(k)_{1-TL}$ with same line parameters. However, we do note that the two uncoupled TLs are still coupled to each other through the electron beam as discussed in Sec. II-C and Fig. 3. Furthermore, when considering coupling between the two TLs, we find even further improvement leading to $\mathrm{Im}(k)_{2-TL} > 2\,\mathrm{Im}(k)_{1-TL}$ in certain regions of Fig. 6.

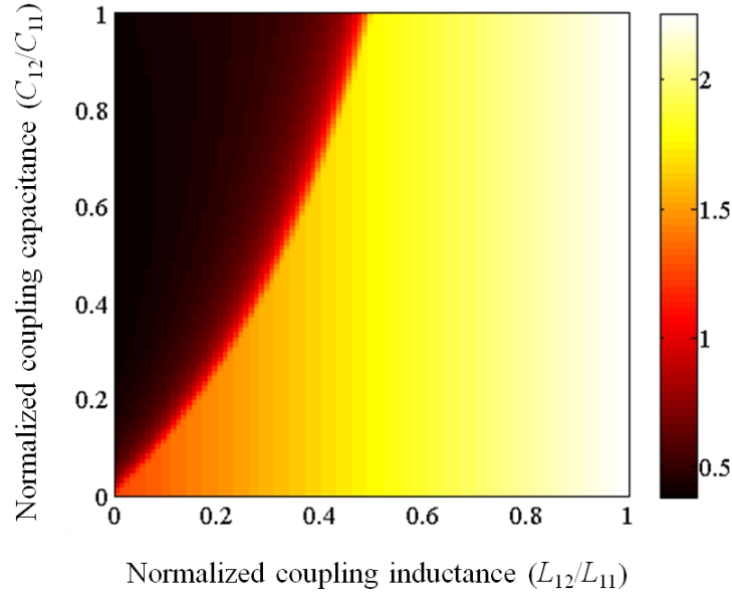

Fig. 6. Plot of ratio of imaginary part of *k* of 2-TL system to imaginary part of *k* of 1-TL system versus the variations in the normalized coupling inductance and capacitance between the two TLs, both coupled to an electron beam.

### III. Power Flow Analysis

Since the beam was included in the active MTL model as voltage-dependent current sources, the power transfer from the beam to the TLs would be equivalent to the power supplied by the dependent current sources to the MTL, which is analyzed next.

#### A. Energy conservation

To show the energy conservation in the active TL system, we define for the $N$ line system, the flux of complex power along the $n^{th}$ TL as $S_{f,n} = \frac{1}{2} V_n I_n^*$, and denote the time-average power flux as $P_{f,n} = \mathrm{Re}(S_{f,n})$. Therefore the total complex power flux in the $N$-TLs at any $z$ is

$$S_f = \sum_{n=1}^{N} S_{f,n} = \frac{1}{2} \mathbf{V}^T \mathbf{I}^*. \tag{19}$$

Consider the variation of complex power along the MTL, $\partial_z S_f = \frac{1}{2}(\partial_z \mathbf{V}^T) \mathbf{I}^* + \frac{1}{2} \mathbf{V}^T (\partial_z \mathbf{I}^*)$. Using the transpose and complex conjugates of (4) and (5), we simplify $\partial_z S_f$ as $\partial_z S_f + \frac{1}{2} j\omega (\mathbf{I}^T \underline{\mathbf{L}} \mathbf{I}^* - \mathbf{V}^T \underline{\mathbf{C}} \mathbf{V}^*) = S_S$, which is interpreted as a one dimensional Poynting theorem with the term $S_S = \frac{1}{2} \mathbf{V}^T \mathbf{i}_S^*$ representing the total complex power delivered per unit length by all the dependent current sources. The term $\partial_z S_f$ is seen as the power flux in the TL along the $+z$ direction whereas the term $j\omega \frac{1}{2}(\mathbf{I}^T \underline{\mathbf{L}} \mathbf{I}^* - \mathbf{V}^T \underline{\mathbf{C}} \mathbf{V}^*)$ is purely imaginary, with $\frac{1}{2} \mathbf{I}^T \underline{\mathbf{L}} \mathbf{I}^*$ and $\frac{1}{2} \mathbf{V}^T \underline{\mathbf{C}} \mathbf{V}^*$ the stored magnetic and electric energy line densities respectively.

#### B. Evaluation of time-average power delivered by dependent current sources

V. A. Tamma and F. Capolino, UC Irvine, May 2013

For the $N$ line system, the time-average power delivered by each dependent current source $i_{S,n}$ is $P_{S,n} = \frac{1}{2}\mathrm{Re}\left(V_n i_{S,n}^*\right)$ and therefore the total time-average power delivered by the $N$ dependent current generators is $P_S = \sum_{n=1}^{N} P_{S,n} = \mathrm{Re}(S_S)$. Using (6), we express $P_S$ in terms of the circuit admittance $\underline{\mathbf{Y}}$ as

$$P_S = \tfrac{1}{2}\mathrm{Re}\left(\mathbf{V}^T \underline{\mathbf{Y}}^* \mathbf{V}^*\right) = \frac{1}{2}\mathrm{Re}\left(-\frac{\left(k^*\right)^2}{j\omega}\mathbf{V}^T \underline{\mathbf{L}}^{-1}\mathbf{V}^*\right), \tag{20}$$

where, we dropped the term $j\omega \mathbf{V}^T \underline{\mathbf{C}} \mathbf{V}^*$ as it is always purely imaginary for any real $\omega$ and hence does not contribute to the time-average power delivered to the TLs. Since $\underline{\mathbf{L}}$ and therefore $\underline{\mathbf{L}}^{-1}$ is a real, symmetric, positive definite matrix, the quadratic form $F(z) = \mathbf{V}^T \underline{\mathbf{L}}^{-1} \mathbf{V}^* = \sum_{n=1}^{N} \lambda_{L,n}^{-1} \left|\mathbf{t}_{I,n}^T \mathbf{V}\right|^2 > 0$, where, $\mathbf{t}_{I,n}$ are the eigenvectors of $\underline{\mathbf{L}}^{-1}$ associated with the eigenvalue $\lambda_{L,n}^{-1}$, and $\lambda_{L,n}$ are the eigenvalues of the matrix $\underline{\mathbf{L}}$. It is convenient to express the voltages in terms of their initial condition as $V_n = V_{n0} e^{j\omega t} e^{-j\beta z} e^{\pm \alpha z}$ such that $\left|\mathbf{t}_{I,n}^T \mathbf{V}\right|^2 = \left|\mathbf{t}_{I,n}^T \mathbf{V}_0\right|^2 e^{\pm 2\alpha z}$ for two conjugate roots $k = \beta \pm j\alpha$, and define $F_0 = \sum_{n=1}^{N} \lambda_{L,n}^{-1} \left|\mathbf{t}_{I,n}^T \mathbf{V}_0\right|^2 > 0$. Then, the time-average power delivered by the dependent current sources in terms of the circuit admittance $\underline{\mathbf{Y}}$ is

$$P_S = \pm \frac{\alpha\beta}{\omega} e^{\pm 2\alpha z} F_0, \tag{21}$$

where, we have used the relation $\mathrm{Re}\left[j\left(k^*\right)^2\right] = \pm 2\alpha\beta$. The power delivered to the TL is different from zero for complex $k = \beta \pm j\alpha$. Using (11), the time-average power delivered by the dependent current sources can also be written in terms of the beam admittance $\underline{\mathbf{Y}}_b$ as $P_S = -\tfrac{1}{2}\mathrm{Re}\left(\mathbf{V}^T \underline{\mathbf{Y}}_b^* \mathbf{V}^*\right) = -\tfrac{1}{2}\mathrm{Re}\left[Y_b^*\left(\mathbf{V}^{*T}\mathbf{sa}^T\mathbf{V}\right)^*\right]$. We identify $\mathbf{V}^{*T}\mathbf{sa}^T\mathbf{V}$ as the quadratic form and observe that $\mathbf{sa}^T$ has only one non-zero eigenvalue $\lambda_1 = \sum_{n=1}^{N}(s_n a_n)$ and has $N-1$ vanishing eigenvalues. The eigenvalue $\lambda_1$ is real since we assumed that both $\mathbf{s}$ and $\mathbf{a}$ are composed of real numbers. Therefore, the time-average delivered power in terms of the beam admittance $\underline{\mathbf{Y}}_b$ is

$$P_S = -\frac{1}{2}\frac{\eta\beta_0\rho_0}{u_0}\lambda_1 \left|\mathbf{q}_1'^T \mathbf{V}\right|^2 \mathrm{Re}\left[\frac{j\left(k^*\right)^2}{\left(\beta_0 - k^*\right)^2}\right], \tag{22}$$

where, $\mathbf{q}_1'$ is the eigenvector of $\mathbf{sa}^T$ associated with the eigenvalue $\lambda_1$. This also shows that time average power is delivered by the electron beam when $k$ is complex, as it will be shown later in more details.

**C. Summary on power transfers for real and complex conjugates $k$**

V. A. Tamma and F. Capolino, UC Irvine, May 2013

The power transferred from the beam to the *N*-coupled TL system is given in (21), which however depends on the complex value of the propagation constant $k$. From Sec. II-E, if (18) is satisfied, we find that $k$ has $2N$ real roots, $k = (\beta_1, \beta_2, \beta_3, ...., \beta_{2N})$ and two conjugate roots $k = \beta \pm j\alpha$. We want to clarify the value and direction of power transferred under these conditions.

Let us consider first the case of the $m^{th}$ real root $k = \beta_m$. Since $\mathbf{V}^T \underline{\mathbf{L}}^{-1} \mathbf{V}^*$ in (20) is purely real and positive, we find $P_S$ to be purely imaginary. Therefore, we have no power transferred between the electron beam and the MTL, i.e., $P_S = 0$.

Next, we investigate the value and direction of time-averaged power transferred for the complex root $k = \beta \pm j\alpha$. In this case we consider the time-average power delivered to the MTL by all the sources per unit length given in (21). Therefore, from (21), if we assume $\beta > 0$, the real power is always transferred from the electron beam to the MTLs in the case of $\alpha > 0$, and vice versa the power is transferred from the MTL to the beam when $\alpha < 0$.

**D. Power flux along the MTL**

For the MTL system the total complex power flux in the *N*-TL system at any $z$ is given by (19) and the time-averaged flux is given by $P_f = \text{Re}(S_f)$. Using (4), (5) and (6) in (19) we obtain,

$$P_f = \tfrac{1}{2}\text{Re}\left(\frac{k^*}{\omega}\mathbf{V}^T \underline{\mathbf{L}}^{-1} \mathbf{V}^*\right) = \tfrac{1}{2}\text{Re}\left(\frac{k}{\omega}\right) e^{\pm 2\alpha z} F_0, \tag{23}$$

which shows that there is always a *positive flux* of power in the MTL along the $+z$ direction for $\text{Re}(k) > 0$. If we take the ratio of $P_S/P_f$ for the complex conjugate wavenumber $k = \beta \pm j\alpha$ we obtain,

$$P_S/P_f = \pm 2\alpha. \tag{24}$$

This means that the ratio of power transferred from the electron beam to the entire MTL system, over the power carried in the MTL system, depends only on the value of $\pm \alpha$.

**E. Analysis of complex power transfer**

From (7) and (10) it is observed that the circuit and beam admittances depend on the complex propagation constant $k = \beta \pm j\alpha$ and therefore are decomposed into real and imaginary components. Accordingly for $k = \beta + j\alpha$, $\underline{\mathbf{Y}} = \underline{\mathbf{Y}}_r + j\underline{\mathbf{Y}}_i$ and $\underline{\mathbf{Y}}_b = \underline{\mathbf{Y}}_{b,r} + j\underline{\mathbf{Y}}_{b,i}$. When higher powers of $\alpha$ were neglected assuming $|\alpha| \ll |\beta|$, the real and imaginary parts can be simplified as $\underline{\mathbf{Y}}_r = \frac{2\beta\alpha}{\omega}\underline{\mathbf{L}}^{-1}$, $\underline{\mathbf{Y}}_i \cong -\frac{\underline{\mathbf{L}}^{-1}}{\omega}\left[\beta^2 \underline{\mathbf{1}} - \omega^2 \underline{\mathbf{L}}\underline{\mathbf{C}}\right]$, and $\underline{\mathbf{Y}}_{b,r} \cong 2\beta_0 \beta \alpha (\beta_0 - \beta) G_0 \mathbf{s}\mathbf{a}^T$, $\underline{\mathbf{Y}}_{b,i} \cong -\left(\beta^2(\beta_0-\beta)^2 - 3\alpha^2 \beta_0^2 + 2\alpha^2 \beta \beta_0\right) G_0 \mathbf{s}\mathbf{a}^T$ with $G_0 = \frac{\beta_0 \eta \rho_0}{u_0} / \left[(\beta_0 - \beta)^2 + \alpha^2\right]^2 > 0$.

The resonance condition (11) implies that $\underline{\mathbf{Y}}$, $\underline{\mathbf{Y}}_b$ and the complex power are related by the quadratic form $\mathbf{V}^T \underline{\mathbf{Y}}^* \mathbf{V}^* + \mathbf{V}^T \underline{\mathbf{Y}}_b^* \mathbf{V}^* = 0$. This identity must be simultaneously verified for real and imaginary power terms, leading to $\mathbf{V}^T \underline{\mathbf{Y}}_r^* \mathbf{V}^* + \mathbf{V}^T \underline{\mathbf{Y}}_{b,r}^* \mathbf{V}^* = 0$ and $\mathbf{V}^T \underline{\mathbf{Y}}_i^* \mathbf{V}^* + \mathbf{V}^T \underline{\mathbf{Y}}_{b,i}^* \mathbf{V}^* = 0$.

V. A. Tamma and F. Capolino, UC Irvine, May 2013

First, considering the real power terms, we can expand $\mathbf{V}^T \underline{\mathbf{Y}}_r^* \mathbf{V}^* + \mathbf{V}^T \underline{\mathbf{Y}}_{b,r}^* \mathbf{V}^* = 0$ as

$$\frac{1}{\omega} F + \beta_0 (\beta_0 - \beta) G_0 \lambda_1 \left| \mathbf{q}_1'^T \mathbf{V} \right|^2 = 0. \tag{25}$$

The first term in (25) is the contribution from $\underline{\mathbf{Y}}_r$ and is positive while the second term is from $\underline{\mathbf{Y}}_{b,r}$ which may assume both signs depending on $(\beta_0 - \beta)$. Therefore, (25) could be satisfied for some $\omega$ only if $\beta > \beta_0$ and $\lambda_1 > 0$ (note that $\lambda_1 > 0$ is also true if $\mathbf{s} = \mathbf{a}$), because under this condition the terms in (25) always have opposite signs. In summary, if $\beta > \beta_0$ and $\alpha > 0$ then $2\beta_0 \beta \alpha (\beta_0 - \beta) < 0$ and thereby the real power $P_S = -\frac{1}{2} \mathrm{Re}\left( \mathbf{V}^T \underline{\mathbf{Y}}_{b,r}^* \mathbf{V}^* \right)$ associated to $\underline{\mathbf{Y}}_{b,r}$ is always positive causing power flow from the beam to the MTL when (11) holds true. Vice versa if $\beta < \beta_0$ and $\alpha > 0$, (25) and so as (11) cannot be satisfied and there are no complex $k$ solutions, implying no power transfer.

Considering the reactive power terms, we can expand $\mathbf{V}^T \underline{\mathbf{Y}}_i^* \mathbf{V}^* + \mathbf{V}^T \underline{\mathbf{Y}}_{b,i}^* \mathbf{V}^* = 0$ as

$$\frac{1}{\omega L_{nn}'} \sum_{n=1}^{N} \left( \beta^2 - \omega^2 L_{nn}' C_{nn}' \right) \left| \mathbf{t}_{I,n}^T \mathbf{V} \right|^2 + \left( \beta^2 (\beta_0 - \beta)^2 - 3\alpha^2 \beta_0^2 + 2\alpha^2 \beta \beta_0 \right) \lambda_1 G_0 \left| \mathbf{q}_1'^T \mathbf{V} \right|^2 = 0, \tag{26}$$

where, $\mathbf{t}_{I,n}^T$ are the eigenvectors of $\underline{\mathbf{Y}}_i$ (that are also eigenvectors of $\underline{\mathbf{L}}^{-1}$) and $L_{nn}^{-1} \left( \beta^2 - \omega^2 L_{nn}' C_{nn}' \right)$ are the eigenvalues of the real symmetric matrix $-\omega \underline{\mathbf{Y}}_i$. Once again, we observe that (26) is a sum of two contributions, the first from $\underline{\mathbf{Y}}_i$ and the second from $\underline{\mathbf{Y}}_{b,i}$, and we look for other conditions to find solutions assuming that $\beta > \beta_0$ (since there are none for $\beta < \beta_0$). It can be shown that the sign of the second term is always negative for $(\beta - \beta_0)^2 < \alpha^2$, which holds true in practical cases. Indeed, for example in the 1-TL case, one finds that $0 < \beta - \beta_0 = \beta_0 C_g / 2 \ll \beta_0$ and $\alpha = \sqrt{3} \beta_0 C_g / 2 > (\beta - \beta_0)$ [3]-[6]. Therefore, we find that the second term in (26) is always negative for $0 < \beta - \beta_0 < \alpha$. This means that if the resonance condition (11) is to be satisfied by a complex $k$, then the first term in (26) should have $\sum_{n=1}^{N} \left( \beta^2 - \omega^2 L_{nn}' C_{nn}' \right) > 0$. Recognizing that $\beta_{c,n}^2 = \omega^2 L_{nn}' C_{nn}'$ (see discussion of (28) in Appendix B for definitions of notation), this means that $\beta \geq \max \{ \beta_{c,1}, \beta_{c,2}, ..., \beta_{c,N} \}$ and $0 < \beta - \beta_0 < \alpha$ ensure that the resonance condition (26) can be always satisfied for any $\omega$. If instead $\beta < \min \{ \beta_{c,1}, \beta_{c,2}, ..., \beta_{c,N} \}$, then, $\sum_{n=1}^{N} \left( \beta^2 - \omega^2 L_{nn}' C_{nn}' \right) < 0$ and the resonance condition (26), and hence (11), will not be satisfied. Therefore, we note that a complex $k$ solution is found when $0 < \beta - \beta_0 < \alpha$, and that the imaginary power $\mathrm{Im}\, S_S = \frac{1}{2} \left( \mathbf{V}^T \underline{\mathbf{Y}}_{b,i}^* \mathbf{V}^* \right)$ delivered *by* the beam (and delivered *to* the MTL) is positive.

To conclude, when $\beta > \beta_0$ and $\alpha > 0$ note that $Y_b$ in Fig. 3 is capacitive since $\mathrm{Im}(Y_b) > 0$, and it has $\mathrm{Re}(Y_b) < 0$, hence it *delivers* positive time-average power to the whole circuit.

**V. Conclusion**

V. A. Tamma and F. Capolino, UC Irvine, May 2013

We have presented an approach to extend the one-dimensional linearized Pierce model to multiple transmission lines coupled to a single electron beam. We have developed two formalisms to evaluate the *k*-ω dispersion relation including a matrix formulation and a formalism developed in terms of admittances. Both formalisms predict that the *N*-TL system coupled to electron beam has 2*N*+2 solutions. Using the derived formalisms, we have predicted conditions which would always guarantee that the system supports growing waves. We have shown that if the electron propagation constant $\beta_0$ is greater than or equal to the largest of the circuit propagation constants $\beta_{c,n}$, then the system always supports a pair of increasing and decreasing waves in addition to *N* of forward and *N* backward waves. The condition is equivalent to stating that the electron velocity should be smaller than the smallest phase velocity among the *N*-TLs. In addition, we have shown that there exist other finite ranges of smaller electron propagation constant $\beta_0$ that also allow for a growing wave solution. These properties have also been verified numerically for some illustrative examples. On the TL side, the beam-line interaction was modeled as a distributed voltage dependant shunt current source. It was found that for the growing wave solution, the beam always supplies power to the transmission lines. We anticipate that the theoretical framework developed in this paper could be used to design high power microwave devices incorporating multiple transmission lines with possible exploitation of degenerate band edges in the Brillouin diagram when using periodicity along *z*.

**Appendix A: reduction of determinant**

The dispersion equation in (13) is derived from $Det[\underline{\mathbf{M}} - k\underline{\mathbf{1}}] = 0$ by use of determinant relation for partitioned matrices, $Det\begin{bmatrix} \underline{\mathbf{A}} & \underline{\mathbf{B}} \\ \underline{\mathbf{C}} & \underline{\mathbf{D}} \end{bmatrix} = Det(\underline{\mathbf{D}}) Det(\underline{\mathbf{A}} - \underline{\mathbf{B}}\underline{\mathbf{D}}^{-1}\underline{\mathbf{C}})$ [26], where, in general, the matrices $\underline{\mathbf{A}}$, $\underline{\mathbf{B}}$, $\underline{\mathbf{C}}$ and $\underline{\mathbf{D}}$, have sizes *M*×*M*, *M*×*L*, *L*×*M* and *L*×*L*, respectively and $\underline{\mathbf{D}}$ is an invertible matrix. In this particular case, the matrices are

$$\underline{\mathbf{A}} = \begin{bmatrix} -k\underline{\mathbf{1}} & \omega\underline{\mathbf{L}} \\ \omega\underline{\mathbf{C}} & -k\underline{\mathbf{1}} \end{bmatrix}, \quad \underline{\mathbf{B}} = \begin{bmatrix} 0 & 0 \\ \omega\eta\dfrac{\rho_0}{u_0^2}\mathbf{s} & -\beta_0 \mathbf{s} \end{bmatrix}, \quad \underline{\mathbf{C}} = \begin{bmatrix} 0 & -\omega(\mathbf{a}^T\underline{\mathbf{L}}) \\ 0 & 0 \end{bmatrix} \quad \text{and} \quad \underline{\mathbf{D}} = \begin{bmatrix} \beta_0 - k & 0 \\ -\omega\eta\dfrac{\rho_0}{u_0^2} & \beta_0 - k \end{bmatrix}.$$

Therefore, $Det[\underline{\mathbf{M}} - k\underline{\mathbf{1}}] = 0$ can be expanded into the form

$$(\beta_0 - k)^2 Det\begin{bmatrix} -k\underline{\mathbf{1}} & \omega\underline{\mathbf{L}} \\ \omega\underline{\mathbf{C}} & -k\underline{\mathbf{1}} - \dfrac{k\omega^2 \eta \rho_0}{u_0^2 (k-\beta_0)^2} \mathbf{s}\mathbf{a}^T \underline{\mathbf{L}} \end{bmatrix} = 0. \tag{27}$$

In (27) we identify that each submatrix is of the same order *N*×*N*. Therefore, using Schur's formulas [26] and after some algebraic manipulation, (27) is reduced to the final form of the dispersion equation in (13).

**Appendix B: growing wave condition**

The matrices $\underline{\mathbf{L}}$, $\underline{\mathbf{C}}$ in (12) can be brought to a diagonal form $\underline{\mathbf{L}}'$, $\underline{\mathbf{C}}'$ using the modal decoupling technique subject to the conditions described in [21]-[23]. We transform (12) to the form (while following the notation in [21]-[23])

$$Det(k^2\underline{\mathbf{1}} - \underline{\boldsymbol{\beta}}_c'^2)\left[1 + \dfrac{\eta\rho_0\beta_0^2 k^2}{(k-\beta_0)^2}\left((\underline{\mathbf{T}}_V^T \mathbf{a})^T \left[(k^2\underline{\mathbf{1}} - \underline{\boldsymbol{\beta}}_c'^2)^{-1} \underline{\mathbf{L}}'\right](\underline{\mathbf{T}}_V^T \mathbf{s})\right)\right] = 0, \tag{28}$$

where, we define $\underline{\boldsymbol{\beta}}_c'^2 = \omega^2 \underline{\mathbf{T}}_V^{-1} \underline{\mathbf{L}} \underline{\mathbf{C}} \underline{\mathbf{T}}_V$ and denote $\beta_{c,n}$ as the positive square root of the $n^{th}$ diagonal element of $\underline{\boldsymbol{\beta}}_c'^2$.

We note that $Det\left(k^2 \underline{\mathbf{1}} - \underline{\boldsymbol{\beta}}_c'^2\right) = \prod_{n=1}^{N}(k^2 - \beta_{c,n}^2)$ and $\left(k^2 \underline{\mathbf{1}} - \underline{\boldsymbol{\beta}}_c'^2\right)^{-1} = \sum_{n=1}^{N}\left(k^2 - \beta_{c,n}^2\right)^{-1}$. When coupled to the electron beam, we assume $k = \beta_0 + \delta$ with $\delta \ll \beta_0$ and therefore $\left(k^2 - \beta_{c,n}^2\right) \approx 2\beta_0\left[\delta + \beta_{c,n}^2/(2\beta_0)\right]$, where we define $\beta_{d,n}^2 = \beta_0^2 - \beta_{c,n}^2$ and assume $\left|\beta_{d,n}^2\right| < \beta_0$ for $n = 1,2,..N$. We assume $\mathbf{s} = \mathbf{a}$ and denote $\tilde{s}_n^2 = \left(\mathbf{t}_{V,n}^T \mathbf{s}\right)^2 > 0$. We note that $\eta\rho_0 = I_0 u_0/(2V_0)$, where $V_0$ is the d.c. beam voltage and $u_0^2 = 2\eta V_0$ [3] such that (28) can be simplified to the form $\delta^2 \prod_{n=1}^{N}\left(\delta + \beta_{d,n}^2/(2\beta_0)\right) + \sum_{m=1}^{N}\left[\tilde{s}_m^2\left(\omega L_{mm}'/(4V_0/I_0)\right)\prod_{\substack{n=1\\n\neq m}}^{N}\left(\delta + \beta_{d,n}^2/(2\beta_0)\right)\right] = 0$, and is a sum of two terms. The first term is a polynomial in $\delta$ of degree $N+2$ with $N+1$ terms and the second term is a sum of $N$ polynomials in $\delta$, with each polynomial of degree $N-1$ and having $N-1$ terms with each polynomial containing the terms $\omega L_{mm}' \delta^n/(4V_0/I_0)$ which are ignored as they are very small since typically $4V_0/I_0$ is a very large number. On expanding and collecting common terms and substituting $M = N+2$, (28) can be written as a polynomial of degree $(N+2)$

$$P(\delta) = +\delta^M + c_{M-1}\delta^{M-1} + .... + c_3\delta^3 + c_2\delta^2 + c_0 = 0. \quad (29)$$

The coefficients $c_{M-1}, c_{M-2},...,c_0$ are formed from various combinations of $\beta_{d,n}^2/(2\beta_0)$ with $c_0 = \sum_{m=1}^{N}\left[\tilde{s}_m^2\left(\omega L_{mm}'/(4V_0/I_0)\right)\prod_{\substack{n=1\\n\neq m}}^{N}\left[\beta_{d,n}^2/(2\beta_0)\right]\right]$, $c_2 = \prod_{n=1}^{N}\left[\beta_{d,n}^2/(2\beta_0)\right]$ and $c_{M-1} = \sum_{n=1}^{N}\left[\beta_{d,n}^2/(2\beta_0)\right]$. We find the following on application of Descartes' sign of rules [28] to (29) and more information can be found in [30].

If $\beta_0 \geq \max\{\beta_{c,1},...,\beta_{c,N}\}$, implies all $\beta_{d,n} > 0$ and hence all coefficients $c_{M-1}, c_{M-2},...,c_0$, are purely real and positive. Hence, (29) has no positive roots, up to $N$ negative roots (or decreased by multiple of 2) and at least $(2,4,....,N,N+2)$ complex conjugate pairs of roots for any $\omega L_{mm}'/(4V_0/I_0) > 0$.

If $\beta_0 < \min\{\beta_{c,1},...,\beta_{c,N}\}$, implies all $\beta_{d,n} < 0$ and hence all coefficients $c_{M-1}, c_{M-2},...,c_0$, are real and negative. Hence, (29) has $N+1$ positive roots (or decreased by multiple of 2), one negative root and at least $(2,4,....,N,N+1)$ complex conjugate pairs of roots for only certain values of $\omega L_{mm}'/(4V_0/I_0) > 0$.

**Appendix C: data for numerical examples**

The following are the parameters used to generate the plots in Figs. 4 and 5. For the 2-TL system whose real and imaginary parts of $k$ are plotted in Fig. 4(a, b), we use $L_{11} = 10$, $L_{22} = 2$, $L_{12} = 0.1$ (all values in nH) and $C_{11} = 12$, $C_{22} = 3$, $C_{12} = 0.2$ (all values in pF). For the 3-TL system whose imaginary part of $k$ is plotted in Fig. 5(a), we use $L_{11} = 5$, $L_{22} = 13$, $L_{33} = 5$, $L_{12} = 0.1$, $L_{13} = 0.5$, $L_{23} = 1$ (all values in nH) and $C_{11} = 14$, $C_{22} = 15$, $C_{33} = 3$, $C_{12} = 1$, $C_{13} = 0.4$, $C_{23} = 10.1$ (all values in pF). For the 3-TL system with all identical lines whose imaginary part of $k$ is plotted in Fig. 5(b), we use self inductance of each line, $L_{nn} = 25$ [nH], n=1,2,3, and the coupling inductance between lines $L_{nm} = 0.01$ [nH], n,m=1,2,3 and $n \neq m$. The self capacitance of each line is $C_{nn} = 2$ [pF], n=1,2,3, and coupling capacitance between lines is $C_{nm} = 0.01$ [pF], n,m=1,2,3 and $n \neq m$. In addition, the inductance and capacitance for the 1-TL used for comparison are $L$=25 [nH] and $C$=2 [pF]. The parameters of the 2-TL system with identical lines used to generate Fig. 6 are: self inductances and self capacitances in the 2-TL are $L_{11} = L_{22} = 10$ [nH] and $C_{11} = C_{22} = 10$ [pF], respectively. The parameters of the 1-TL system used for comparison in Fig. 6 are: self inductance and self capacitance of the TL are $L = 10$ [nH] and $C = 10$ [pF], respectively. In all cases, we assume $\mathbf{s} = \mathbf{a}$ with $s_n = 1, \forall n$. The frequency of operation in all calculations was assumed to be 10 GHz and the magnitude of line charge density is kept constant at $\rho_0 = 0.377$ [nC/m].

**Acknowledgements**

This research was supported by AFOSR MURI Grant FA9550-12-1-0489 administered through the University of New Mexico. The authors are grateful to Alexander Figotin and Guillermo Reyes from the Department of Mathematics, University of California, Irvine for helpful discussions.